\documentclass[aps,prl,twocolumn,final,letterpaper]{revtex4}

\usepackage{graphicx}   
\usepackage{import}                         
\usepackage{epstopdf}
\usepackage{amsmath} 
\usepackage{bm}
\usepackage{amssymb}
\usepackage{quotes}
\usepackage{transparent}
\usepackage{dcolumn}
\usepackage[usenames, dvipsnames]{color}
\usepackage{multirow}
\usepackage{cancel} 
\usepackage{mdframed}
\usepackage{color}
\usepackage{bm}
\usepackage{setspace}
\usepackage{dsfont}
\usepackage{braket}
\usepackage{slashed}
\usepackage{float}
\usepackage{soul, color} 
\soulregister\ref{7}  
\soulregister\cite{7} 
\renewcommand{\st}[1]{}

\newcommand{\weg}{\omega_{eg}}
\begin{document}
\rmfamily

\title{Extreme enhancement of spin relaxation mediated by surface magnon polaritons}
\author{Jamison Sloan$^{1*\dagger}$, Nicholas Rivera$^{1*}$, John D. Joannopoulos$^{1}$, Ido Kaminer$^{2}$, and Marin Solja\v{c}i\'{c}$^{1}$}

\affiliation{$^{1}$ Department of Physics, MIT, Cambridge, MA 02139, USA \\
$^{2}$ Department of Electrical Engineering, Technion $-$ Israel Institute of Technology, Haifa 32000, Israel. \\
$\dagger$ Corresponding author e-mail: jamison@mit.edu \\
* These authors contributed equally to this work}

\noindent	

\begin{abstract}
Polaritons in metals, semimetals, semiconductors, and polar insulators, with their extreme confinement of electromagnetic energy, provide many promising opportunities for enhancing typically weak light-matter interactions such as multipolar radiation, multiphoton spontaneous emission, Raman scattering, and material nonlinearities. These highly confined polaritons are quasi-electrostatic in nature, with most of their energy residing in the electric field. As a result, these "electric" polaritons are far from optimized for enhancing emission of a magnetic nature, such as spin relaxation, which is typically many orders of magnitude slower than corresponding electric decays. Here, we propose using surface magnon polaritons in negative magnetic permeability materials such as MnF$_2$ and FeF$_2$ to strongly enhance spin-relaxation in nearby emitters in the THz spectral range. We find that these magnetic polaritons in 100 nm thin-films can be confined to lengths over 10,000 times smaller than the wavelength of a photon at the same frequency, allowing for a surprising twelve orders of magnitude enhancement in magnetic dipole transitions. This takes THz spin-flip transitions, which normally occur at timescales on the order of a year, and forces them to occur at sub-ms timescales. Our results suggest an interesting platform for polaritonics at THz frequencies, and more broadly, a new way to use polaritons to control light-matter interactions.
\end{abstract}

\maketitle

\noindent

Polaritons, collective excitations of light and matter, offer the ability to concentrate electromagnetic energy down to volumes far below that of a photon in free space \cite{atwater2007promise,Basov:2016,basov2017towards,low2017polaritons,iranzo2018probing,ni2018fundamental}, holding promise to achieve the long-standing goal of low-loss confinement of electromagnetic energy at the near-atomic scale. The most famous examples are surface plasmon polaritons on conductors, which arise from the coherent sloshing of surface charges accompanied by an evanescent electromagnetic field. These collective excitations are so widespread in optics that their manipulation is referred to as ``plasmonics.'' Plasmons enjoy a myriad of applications, particularly in spectroscopy due to their enhanced interactions with matter. This enhancement applies to spontaneous emission, Raman scattering, optical nonlinearities, and even dipole-``forbidden'' transitions in emitters \cite{moskovits1985surface,albrecht1977anomalously,jeanmaire1977surface,nie1997probing,kauranen2012nonlinear,andersen2011strongly,takase2013selection,rivera2016shrinking}. Beyond plasmons in metals, polaritons in polar dielectrics, such as phonon polaritons \cite{caldwell2013low,caldwell2014sub,dai2014tunable,caldwell2015low} are now being exploited for similar applications due to their ability to concentrate electromagnetic energy on the nanoscale in the mid-IR/THz spectral range.

\begin{figure*}[ht]
	\centering
	\includegraphics[width=\linewidth]{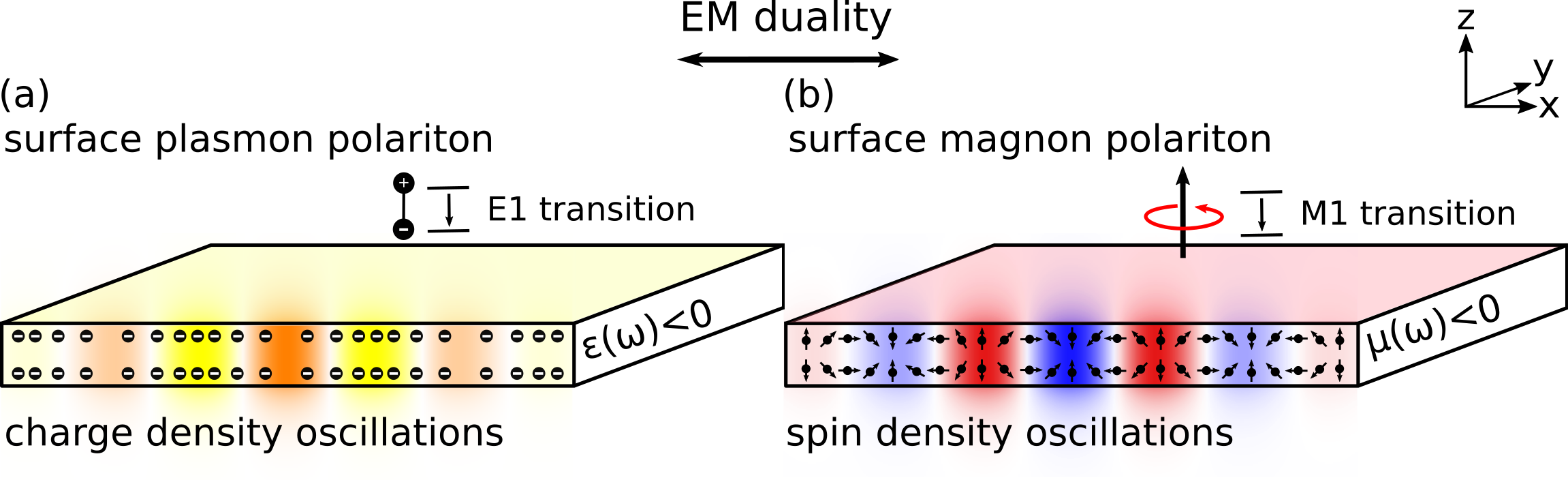}
	\caption{\textbf{Electromagnetically dual relationship between surface plasmon polaritons on negative permittivity materials and surface magnon polaritons on negative permeability materials.} (a) Surface plasmon polariton represented as charge density oscillations in a negative $\epsilon$ material. These quantum fluctuations can couple strongly to an electric dipole emitter near the surface to drive enhanced spontaneous emission. (b) Surface magnon polariton represented as a spin density oscillation in a negative $\mu$ material. These quantum fluctuations can couple strongly to a magnetic dipole emitter near the surface to drive enhanced spontaneous emission. Both electric and magnetic surface polaritons can exhibit strong mode confinement, helping to overcome the mismatch between mode wavelength and emitter size.}
	\label{fig:schematic}
\end{figure*}

The ability of nano-confined polaritons to strongly enhance electromagnetic interactions with matter can ultimately be understood in terms of electromagnetic energy density. An electromagnetic quantum of energy $\hbar\omega$, confined to a volume $V$, leads to a characteristic root-mean-square electric field of order $\sqrt{\frac{\hbar\omega}{\epsilon_0 V}}$.  In the case of field interaction with an electron in an emitter, this characteristic field drives spontaneous emission, and thus concentration of energy to smaller volumes leads to enhanced emission. This well-studied phenomenon is best known as the Purcell effect \cite{purcell1946purcell}. Interestingly, if one looks at the electromagnetic energy distribution of a highly confined plasmon- or phonon- polariton, one finds that an overwhelming majority of this energy resides in the electric field. For a polariton with a wavelength 100 times smaller than that of a photon at the same frequency, the energy residing in the magnetic field is of the order of a mere $0.01\%$ of the total energy $\hbar\omega$. This largely suggests that such excitations are relatively inefficient for enhancing spontaneous emission processes which couple to the magnetic field, such as spin-flip transitions or magnetic multipole decays. Nevertheless, enabling magnetic decays at very fast rates represents a rewarding challenge, as increasing rates of spontaneous emission can provide new opportunities for detectors, devices, and sources of light. The Purcell enhancement of magnetic dipole transitions has been approached by a few basic means: the use of highly confined resonances at optical frequencies \cite{rolly2012promoting,hussain2015enhancing}, metamaterials \cite{slobozhanyuk2014magnetic, mahmoud2014wave} and for microwave frequencies, materials with simultaneously very high quality factor and highly confined fields. These advances are reviewed in Ref. \cite{baranov2017modifying}. Many of these methods have the benefit of compatibility with well known materials and use at optical frequencies, 
but the Purcell enhancements in these cases are typically very far from maximal Purcell enhancements that can be achieved with "electric" polaritons at similar frequencies \cite{koppens2011graphene,kumar2015tunable,rivera2016shrinking,miller2017limits,rivera2017making,kurman2018control}. This prompts the question: what kind of electromagnetic response allows one to achieve a similar degree of very strong enhancement for magnetic decays?


The duality between electric and magnetic phenomena, combined with ideas from plasmonics and nano-optics, suggests a new pathway for achieving strong magnetic transition enhancement: highly confined magnetic modes in materials with negative magnetic \emph{permeability}. In particular, plasmon- and phonon-polaritons are associated with a negative dielectric permittivity $\epsilon(\omega)$. By electromagnetic duality, if one replaces $\epsilon(\omega)$ with the magnetic permeability $\mu(\omega)$, then the electric field $\mathbf{E}$ in the dielectric structure becomes the magnetic field $\mathbf{H}$ in the dual magnetic structure. Thus, to very efficiently enhance magnetic decays, one desires a material with negative $\mu(\omega)$ which supports modes dual to "electric" surface polaritons. While likely not the only example, antiferromagnetic resonance is a well-studied example of a phenomenon which can provide precisely this permeability, and the corresponding modes are surface magnon polaritons. 

Here, we propose enhancement of spin relaxation in emitters using highly confined magnon polaritons. We show that the interaction of a magnetic dipole with modes that are primarily magnetic in nature can level the playing field between electric and magnetic processes. Specifically, we find that these systems can shrink the wavelength of light by factors over 10,000, and predict speedups of magnetic dipole spontaneous emission processes on the order of $10^{12}$. Such enhancements could enable extremely slow magnetic decays with radiative lifetimes on the order of a year to occur at sub-millisecond timescales.

The organization of this manuscript is as follows: in section I, we review the electrodynamics of surface magnon polaritons, and derive the dispersion relation and mode profile of magnon polariton modes for the example of an antiferromagnet. In section II, we develop the theory of spin relaxation of emitters into these modes, and in section III, we provide quantitative results for the spontaneous emission by spin systems near existing magnon-polaritonic materials, such as MnF$_2$ and FeF$_2$.

\section{Surface magnon polariton modes}

We begin by reviewing the confined modes which exist on thin films of materials with negative permeability. The modes we describe are surface magnon polaritons \cite{mills1974polaritons,almeida1988dynamical,camley1982surface,shu1982surface} with $\text{Re}\,\mu(\omega) \leq 0$. For the specific case of an antiferromagnetic material near resonance, the frequency-dependent permeability takes the form of a Lorentz oscillator which depends on the microscopic magnetic properties of the antiferromagnetic crystal. Studies of the crystal structures of important antiferromagnetic materials can be found in \cite{stout1954crystal}. The magnetic permeability function for antiferromagnetic resonance (AFMR) derived in \cite{kittel1951theory} is
\begin{equation}
	\mu_{xx} = \mu_{yy} = 1 + \frac{2\gamma^2 H_A H_M}{\omega_0^2 - (\omega + i\Gamma)^2},
    \label{eq:permeability_model}
\end{equation}
with coordinates shown in Figure~\ref{fig:schematic}. In Equation~\ref{eq:permeability_model}, $\omega_0$ is the resonance frequency, $H_A$ is the anisotropy field, $H_M$ is the magnetization field, $\gamma$ is the gyromagnetic ratio, and $\Gamma = 1/\tau$ is a phenomenological damping parameter inversely proportional to the loss relaxation time $\tau$. Furthermore, in the approximation of low damping, the resonant frequency is given as $\omega_0 = \gamma\sqrt{2H_A(H_A + H_E)}$, where $H_E$ is the exchange field which is representative of the magnetic field required to invert neighboring spin pairs. For antiferromagnetic materials such as MnF$_2$ and FeF$_2$, the resonance frequencies $\omega$ takes values 1.69$\times 10^{12}$ and 9.89$\times 10^{12}$ rad/s respectively, and have negative permeability over a relatively narrow bandwidth on the scale of a few GHz. Most importantly for our purposes, $\text{Re}\mu(\omega) < 0$ for $\omega < \omega_0 < \omega_{\text{max}}$, which will permit surface-confined modes. Table~\ref{tab:parameter_table} shows values of material parameters for a variety of antiferromagnetic materials.

\begin{table}[h!]
	\centering
	\begin{tabular}{|l|l|l|l|l|l|} \hline
		Material & $H_A (T)$ & $H_E$ (T) & $M_S$ (T)& $\omega_0$ ($\times 10^{12}$ rad/s) & $\tau$ (s)\\ \hline
		MnF$_2$ & 0.787 & 53.0 & 0.06 & 1.69 & $7.58\times 10^{-9}$ \\ \hline
        FeF$_2$ & 19.745 & 53.3 & 0.056 & 9.89 & $1.06\times 10^{-10}$ \\ \hline
        GdAlO$_3$ & 0.365 & 1.88 & 0.062 & 0.23 & -- \\ \hline
	\end{tabular}
	\caption{Anisotropy fields, exchange fields, sublattice magnetization, resonance frequencies, and damping constants (where known) for antiferromagnetic materials that can support SMPs. Parameters are taken from Refs. \cite{dumelow1997continuum, luthi1983surface}. }
	\label{tab:parameter_table}
\end{table}

\begin{figure*}[ht]
	\centering
	\includegraphics[width=\linewidth]{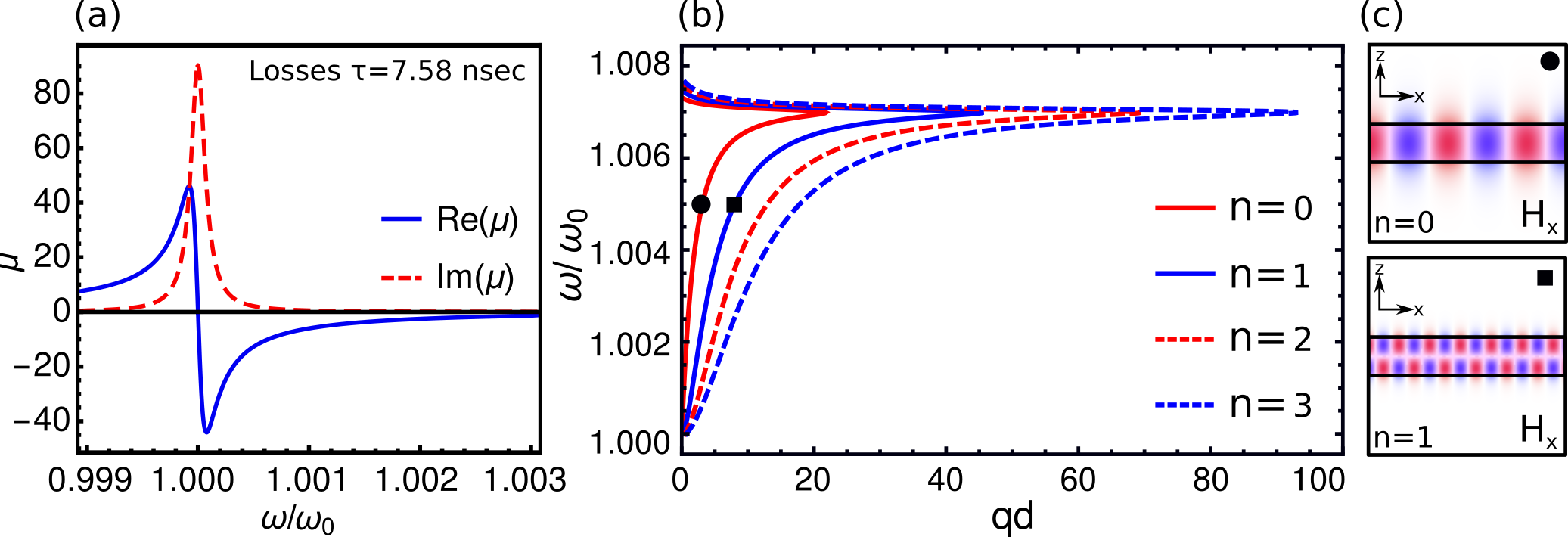}
	\caption{\textbf{Surface magnon polariton (SMP) modes on MnF$_2$.} (a) Frequency dependent permeability function for MnF$_2$ calculated using Equation~\ref{eq:permeability_model} and using the parameters given in Table~\ref{tab:parameter_table}. For MnF$_2$, the resonance frequency is $\omega_0 = 1.68 \times 10^{12}$ rad/s. For $\omega_0 < \omega < \omega_{\text{max}}$, $\text{Re}(\mu) < 0$, allowing for surface modes. (b) Dispersion relation for MnF$_2$ of thickness $d$, calculated in the quasi-magnetostatic limit which is valid in the range of thicknesses $d$ we consider. The first four modes are shown. The inset plot shows the confinement factor $\eta = qc/\omega$ for the first two modes as a function of frequency. (c) Visualization of fundamental and first harmonic mode SMP through the scalar magnetic potential $\psi_H$ shown for a $d=200$ nm film of MnF$_2$ at $\omega/\omega_0 = 1.005$. The locations of these two modes are indicated on the dispersion curve.}
    	\label{fig:dispersion}
\end{figure*}

Antiferromagnetic fluorides exhibit a uniaxial permeability structure with two orthogonal components of the permeability tensor given by $\mu(\omega)$ above, and the other orthogonal component as unity. We start by focusing on crystal orientations in which $\mu = (\mu(\omega), \mu(\omega), 1)$. It is also worthwhile to note that experiments, specifically on nonreciprocal optical phenomena \cite{remer1986nonreciprocal}, have  been performed on these materials in a less conventional geometry where $\mu = (\mu(\omega), 1, \mu(\omega))$. The in-plane anisotropy of this configuration substantially complicates the dispersion relation and propagation structure of the modes. As such, we focus primarily on the former case, but present results for the latter at the end of the text.

For concreteness, we focus on MnF$_2$, a material which has been studied in depth both in theory and experiment \cite{greene1965observation,stamps1991nonreciprocal}, and also exhibits a relatively low propagation loss. We note that FeF$_2$ is also a promising candidate with higher resonance frequency, but also higher loss \cite{hutchings1970spin,brown1994nonreciprocal}. We solve for surface magnon polaritons supported by optically very thin (here, sub-micron) MnF$_2$ films surrounded by air. For the confined modes we consider, the effect of retardation is negligible, and thus we can find the magnon modes using a quasi-magnetostatic treatment \cite{camley1980long}. In the absence of retardation, the electric field is negligible, and the magnetic field, since there are no free currents, satisfies $\nabla\times\mathbf{H} = 0$. Thus the magnetic field can be written as the gradient of a scalar potential $\mathbf{H} = \nabla\psi_H$. This scalar potential then satisfies a scalar Laplace equation
\begin{equation}
    \partial_i \mu_{ij}(\omega) \partial_j \psi_H = 0,
\end{equation}
where we have used repeated indices to denote summation. Applying boundary conditions for continuity of the magnetic potential at the two interfaces of a film of thickness $d$ gives the dispersion relation
\begin{equation}
	q_n = \frac{1}{d\sqrt{-\mu(\omega)}}\left[\tan^{-1}\left(\frac{1}{\sqrt{-\mu(\omega)}}\right) + \frac{n\pi}{2}\right],
\end{equation}
where $n$ is an integer, $q_n$ is the in-plane wavevector of mode $n$, and $\mu(\omega)$ is the  permeability given in Equation~\ref{eq:permeability_model}. We see that $q_n$ is inversely proportional to the thickness of the slab $d$. Identically to confined modes on thin films of plasmonic materials (silver and gold for instance), a thinner film results in a smaller wavelength. Figure~\ref{fig:dispersion}c shows plots of the scalar potential $\psi_H$ associated with SMP modes on MnF$_2$, which is proportional to the magnetic field in direction of propagation. The scalar potential solutions to the Laplace equation take the form  
\begin{equation}
\psi_H^{n}(\mathbf{r},\omega) = \begin{cases}
	e^{i\mathbf{q}_n\cdot\rm{\rho}}e^{-q_n|z|} & |z| > d/2 \\
    \left(\frac{e^{-q_nd}}{f(q_nd)}\right) e^{i\mathbf{q}_n\cdot\rm{\rho}}f(q_nz) & |z| < d/2
\end{cases},
\end{equation}
where $\rho = (x,y)$ is the in-plane position, $f(x) = \cos(x)$ for even modes, and $f(x) = \sin(x)$ for odd modes. Taking the gradient of the scalar potential gives the fully vectorial magnetic field, which reveals that the surface magnon polariton mode propagates in the in-plane direction $\hat{q}$ with circular polarization $\hat{\varepsilon}_{\mathbf{q}} = (\hat{q} + i\hat{z})/\sqrt{2}$. This polarization is well known to be typical of quasistatic surface polariton modes, whether they are the transverse magnetic modes associated with quasielectrostatic excitations or transverse electric modes associated with quasimagnetostatic excitations.

\begin{figure*}[ht]
	\centering
	\includegraphics[width=\linewidth]{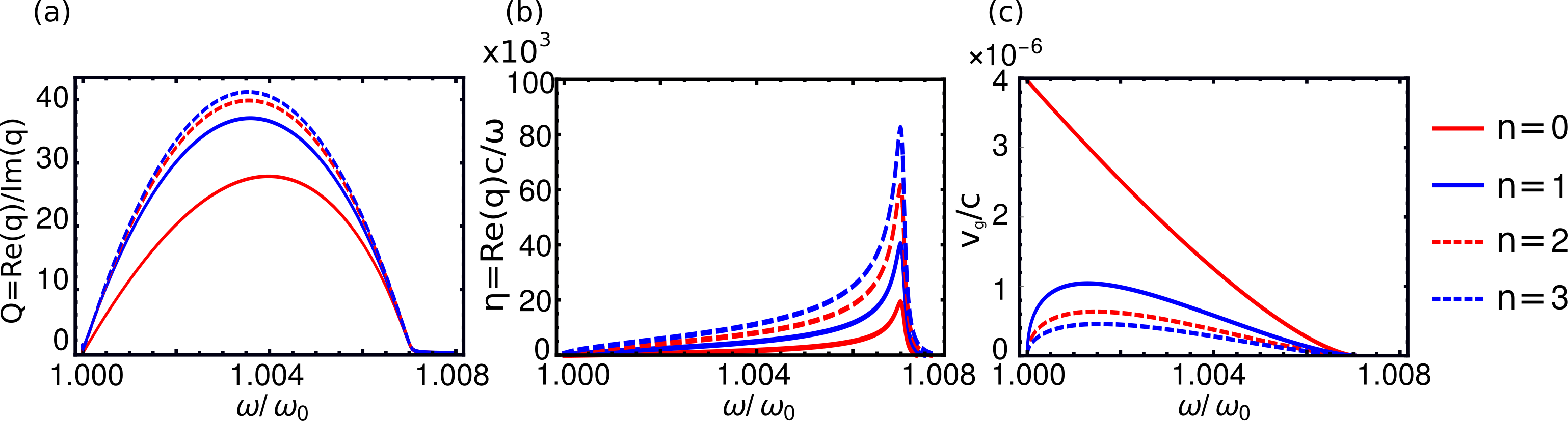}
	\caption{\textbf{Propagation properties of SMP modes on MnF$_2$.} The following dimensionless quantities are plotted for MnF$_2$ with propagation loss $\tau = 7.58$ nsec for the first 4 modes indexed by $n=(0,1,2,3)$. (a) Mode quality factor $Q = \text{Re}(q)/\text{Im}(q)$ as a function of mode frequency. (b) Mode confinement factor $\eta=qc/\omega$ as a function of mode frequency. (c) Normalized group velocity $v_g/c = (1/c)|d\omega/dk|$ as a function of mode frequency.}
    	\label{fig:properties}
\end{figure*}

In Figure~\ref{fig:dispersion}b, we plot the material-thickness-invariant dispersion relation $\omega(qd)$. The dimensionless wavevector $qd$ indicates how the size of the in-plane wavevector compares to the thickness of the film. The dispersion plot shows the first four bands -- the fundamental mode as well as three higher harmonics. Due to the the reflection symmetry of the geometry in the $z$-direction, two of these modes are even parity, and two are odd parity. We can interpret the mode index as the number of half oscillations which the magnetic field makes in the $z$-direction of the film. Higher order modes will have larger wavevectors. Once again, we can further understand the dispersion relation of these modes through analogy to existing polaritonic systems. Specifically, MnF$_2$ is a hyperbolic material since $\mu_\perp > 0$ while $\mu_\parallel < 0$ (where the directions $\perp$ and $\parallel$ are taken with respect to the $z$ axis). This is much like the naturally occurring hyperbolic material hexagonal boron nitride, which has one component of its permittivity negative, while another component is positive. As a result of this, these systems have a multiply-branched dispersion, and the electromagnetic fields are guided inside the crystal. 

The most impressive figure of merit of these modes is the size of their wavelength in comparison to the free space wavelength at a given frequency, also known as a confinement factor or effective index of the mode. Figure~\ref{fig:properties}b highlights this, showing the confinement factor $\eta = qc/\omega = \lambda_0/\lambda_{\text{SMP}}$ for the first four modes ($n=0,1,2,3$) on $d=200$ nm MnF$_2$ as a function of frequency. We see that the fundamental mode reaches a peak confinement of $\eta = 2\times 10^4$, while the first harmonic is confined to twice that with $\eta = 4\times 10^4$.

These values exceed by nearly two orders of magnitude the maximum confinement values that have been observed in common plasmonic media such as thin films of silver, gold, or titanium nitride, or doped graphene. Furthermore, since the confinement scales linearly with $q\sim 1/d$, decreasing the material thickness increases the achievable range of confinement factors. As a simple example of this, consider that a material thickness of $d=50$ nm would correspond to a wavevector 4 times larger than for $d=200$ nm, in other words a maximum fundamental mode confinement of $8\times 10^4$, and a confinement above $10^4$ for much of the surface magnon band.

An explanation for this  high confinement in terms of most basic principles is that the frequencies at which surface magnon polaritons exist (GHz-THz) are orders of magnitude lower than for plasmons which typically exist in IR to optical regimes. Simultaneously, the scale of the wavevector $q$ in both plasmonic and magnonic media is set by the film thickness $d$ (such that plasmons and magnons will have similar wavevectors). In other words, at a fixed material thickness, lower frequency surface magnons have substantially higher potential for geometrical squeezing than surface plasmons. We note that this is not of purely formal interest, as when considering the enhancement of spontaneous emission, one finds that the enhancement is proportional to a power of precisely this confinement factor.

In addition to understanding the confinement of magnon polaritons, it is also important to understand their propagation characteristics, such as propagation quality factor, and group velocity. Figure~\ref{fig:properties}a,c shows the quality factor $Q = \text{Re}(q)/\text{Im}(q)$, as well as the normalized group velocity $v_g/c$ as a function of frequency for the first four modes. We see that propagation losses are lowest toward the middle of the allowed frequency band, showing quality factors greater than 20 for $n=0$. Additionally, we see that the group velocity $v_g$ reaches its maximum near the lower portion of the allowed frequency range, and goes toward zero at the other end.

\section{Theory of spin relaxation into magnon polaritons}


We now discuss the mechanisms that can allow an emitter to couple to highly confined SMPs, and then calculate rates of emitter relaxation associated with SMP emission. A magnetic field can couple to both the electron spin angular momentum and orbital angular momentum, as both angular momenta contribute to the electron's magnetic moment. We describe this interaction quantum mechanically with a interaction Hamiltonian $H_{\text{int}}$ between an emitter and a magnetic field
\begin{equation}
	H_{\text{int}} = -\bm{\mu}\cdot\mathbf{B} = -\frac{\mu_B(\mathbf{L} + g\mathbf{S})}{\hbar}\cdot\mathbf{B},
    \label{eq:hamiltonian}
\end{equation}
where $\bm{\mu}$ is the total magnetic moment of the atom, $\mathbf{S} = \hbar\bm{\sigma}$ is the spin angular momentum operator, $\mathbf{L}$ is the orbital angular momentum operator, $g \approx 2.002$ is the Land\'{e} g-factor. In this Hamiltonian, we note that $\mathbf{B}$ is the quantized magnetic field operator associated with SMP modes.

In order to provide a fully quantum mechanical description of the interactions, we use the formalism of macroscopic QED to express the magnetic field operator as a mode expansion over SMP modes. This approach is similar to that in \cite{glauber1991quantum}, which was applied to quantize electromagnetic fields in dielectric structures. We consider a geometry of a negative $\mu$ material which is translation invariant (i.e., a slab geometry). In this case, the modes are labeled by an in-plane wavevector $\mathbf{q}$. We find then that the magnetic field Schrodinger operator at time $t=0$ takes the form:
\begin{equation}
	\mathbf{B}(\mathbf{r}) = \sum_{\mathbf{q}}\sqrt{\frac{\mu_0\hbar\omega}{2 A C_q}}\left(\hat{\varepsilon}_{\mathbf{q}}e^{i \mathbf{q}\cdot\mathbf{\rho}}e^{-qz}a_{\mathbf{q}} + \hat{\varepsilon}_{\mathbf{q}}^*e^{-i \mathbf{q}\cdot\mathbf{\rho}}e^{-qz}a^\dagger_{\mathbf{q}}\right).
    \label{eq:bfield_operator}
\end{equation}
where $a_{\mathbf{q}}^\dagger$ and $a_{\mathbf{q}}$ are creation and annihilation operators for the SMP modes satisfying the canonical commutation relation $[a_{\mathbf{q}},a_{\mathbf{q}'}^\dagger] = \delta_{\mathbf{qq}'}$, $\hat{\varepsilon}_{\mathbf{q}}$ is the mode polarization, $A$ is the area normalization factor, and $C_q = \int dz\,\mathbf{H}^*(z)\cdot \frac{d(\mu\omega)}{d\omega}\cdot\mathbf{H}(z)$ is a normalization factor ensuring that the mode $\mathbf{H} = \nabla\psi_H$ has an energy of $\hbar\omega_{\mathbf{q}}$. The energy has been calculated according to the Brillouin formula for the electromagnetic field energy in a dispersive medium in a transparency window \cite{landau2013electrodynamics, archambault2010quantum}. In this expression for the energy, we have also used the fact that the modes are magnetostatic in nature, and that the electric energy associated with them is negligible. 

To understand the strength of the coupling between an emitter's spin and SMPs, we calculate spontaneous emission of a spin into a thin negative $\mu$ material such as an antiferromagnet, using Fermi's golden rule. The rate of spin relaxation by emission of a magnon of wavevector $\mathbf{q}$ is given as 
\begin{equation}
	\Gamma_{\mathbf{q}}^{(eg)} = \frac{2\pi}{\hbar^2}|\braket{g, \mathbf{q}|H_{\text{int}}|e,0}|^2\delta(\omega_{\mathbf{q}} - \weg)
    \label{eq:fermi_golden_rule}
\end{equation}

We specify the initial and final states of the system as $\ket{e,0}$ and $\ket{g,\mathbf{q}}$ respectively, where $e$ and $g$ index the excited and ground states of the emitter, $\mathbf{q}$ is the wavevector of the magnon resulting from spontaneous emission, $\omega_{\mathbf{q}}$ is its corresponding frequency, and $\weg$ is the frequency of the spin transition.

Substituting Equation~\ref{eq:bfield_operator} into the Hamiltonian of Equation~\ref{eq:hamiltonian}, and then applying Fermi's golden rule as written in Equation~\ref{eq:fermi_golden_rule}, we find that the spontaneous emission rate $\Gamma^{(eg)}$ per unit magnon in-plane propagation angle $\theta$ is given by:
\begin{equation}
	\frac{d\Gamma^{(eg)}_{\text{dipole}}}{d\theta} = \frac{\mu_B^2\mu_0\weg}{2\pi\hbar}\frac{q^3(\weg)}{C_q(\weg)|v_g(\weg)|}e^{-2q(\weg)z_0}|M_{eg}|^2
    \label{eq:dipole_transition_rate},
\end{equation}
where $|v_g| = |\nabla_{\mathbf{q}}\omega|$ is the magnitude of the SMP group velocity and $M_{eg} = \braket{g|\hat{\epsilon}_{\mathbf{q}}\cdot(\mathbf{L} + g\mathbf{S})|e}$ is the matrix element which describes the transition. In cases where the transition corresponds only to a change of spin of the electron, this matrix element is simply proportional to $\bm{\sigma}_{eg} = \braket{\downarrow|\bm{\sigma}\cdot\hat{\varepsilon}_{\mathbf{q}}|\uparrow}$. Here, the angular dependence can come solely from the magnon polarization.  For a spin transition oriented along the $z$ (ie. out-of-plane) axis, the transition strength into modes at different $\theta$ will be the same, and thus the distribution of emitted magnons isotropic. Spin transitions along a different axis will break this symmetry, resulting in angle dependent emission. In any case, the total rate of emission is obtained by integrating over all angles as $\Gamma^{(eg)}_{\text{dipole}} = \int_0^{2\pi} \left(\frac{d\Gamma^{(eg)}}{d\theta}\right)\,d\theta$.

This formalism can be extended to include losses using the methodology established in \cite{scheel2008macroscopic}. It was found explicitly in \cite{rivera2016shrinking} that in general the presence of losses does not drastically change the total decay rate of the emitter, unless the emitter is at distances from the material much smaller than the inverse wavevector of the modes that are emitted. In the particular case of SMPs of MnF$_2$, the modes have quality factors of $Q \sim 20-30$, and the distances chosen are fairly large, so neglecting material losses is justified. Having presented the general framework for analyzing SMP emission, we now present specific results for SMP emission into a thin film of MnF$_2$.

\section{Transition Rate Results}

\subsection{Magnetic Dipole Transition Rates}

We first discuss the transition rates and associated Purcell factors of magnetic dipole emitters. For a $z$-oriented spin flip of frequency $\weg$ placed a distance $z_0$ from the surface of a negative $\mu$ film, the spontaneous emission rate is given as 
\begin{equation}
	\Gamma_{\text{dipole}}^{(eg)} = \frac{\mu_B^2 \mu_0 \weg}{\hbar} \frac{q^2(\weg)}{C'(\weg)|v_g(\weg)|}e^{-2q(\weg)z_0},
    \label{eq:z_dipole_transition_rate}
\end{equation}
where $C'(\omega) = C(\omega)/q(\omega)$ is a quantity is introduced to remove the wavevector dependence from the normalization. We also note that the group velocity $|v_g(\omega)| \propto 1/q(\omega)$, and thus the whole expression carries a wavevector dependence of $\Gamma_{\text{dipole}}^{(eg)} \propto q^3(\weg)$. 

\begin{figure}[h!]
	\centering
	\includegraphics[width=\linewidth]{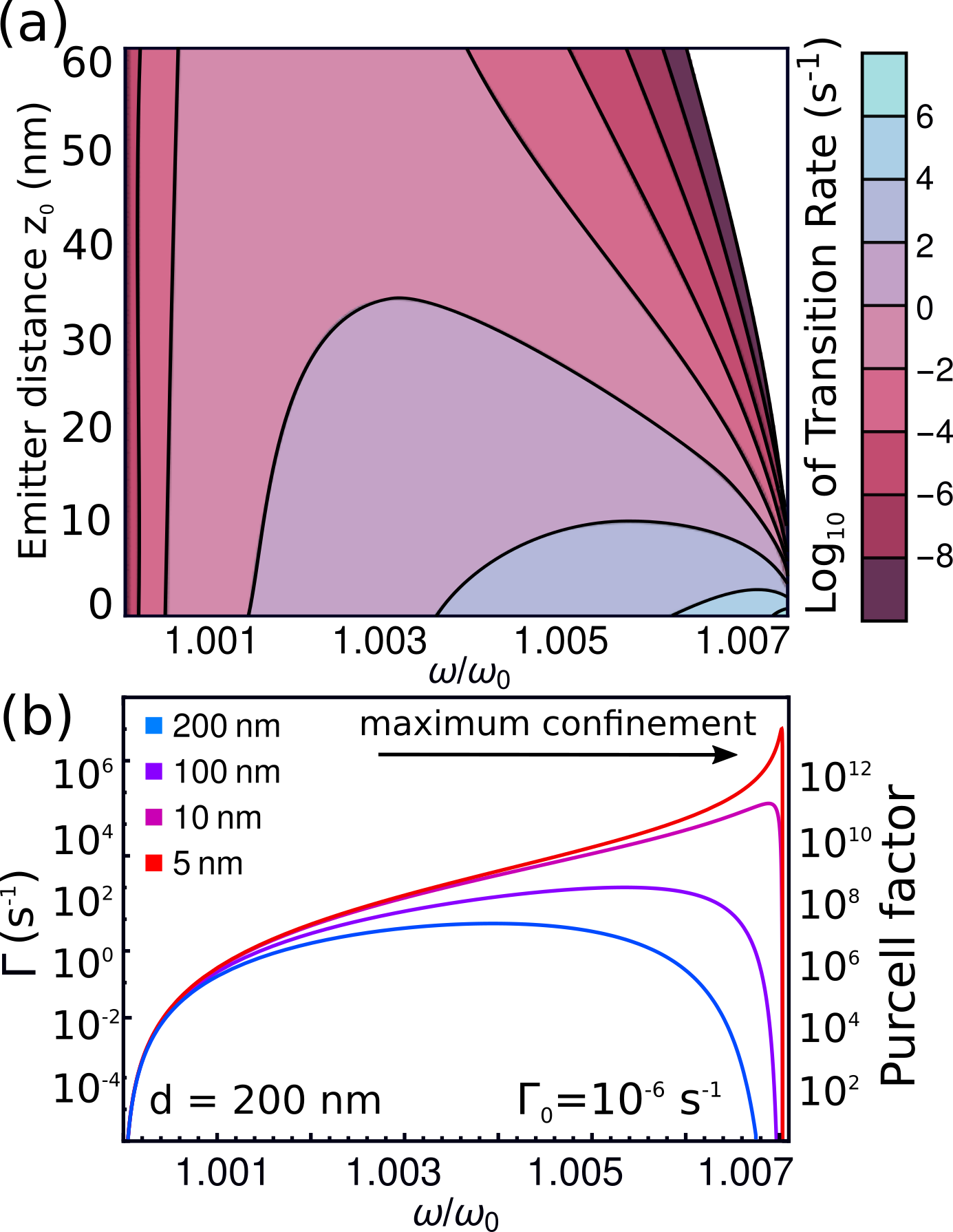}
	\caption{\textbf{Dipole transition rate enhancement by SMPs.} (a) Dipole transition rate for a $z$-oriented spin flip as a function of normalized frequency and distance $z_0$ from the emitter to the surface of a $d=200$ nm MnF$_2$ film. The transition rates decay exponentially with increasing distance from the surface.  (b) Line cuts of the information shown in (a) for different fixed distances $z_0$. The axis on the left shows the total transition rate, while the axis on the right shows the Purcell factor, in other words, the transition rate normalized by the free space transition rate.}
	\label{fig:dipole_rates}
\end{figure}

Figure~\ref{fig:dipole_rates} shows the emission rate as a function of frequency $\omega$ and emitter distance $z_0$ for a $d=200$ nm MnF$_2$ film. Panel (b) shows line cuts of the dipole transition rate at various emitter distances $z_0$.  In this geometry we find that for the highest supported magnon frequencies, the total rate of emission may exceed $10^5$ s$^{-1}$, which corresponds to a decay time of 10 $\mu$s. This is eleven orders of magnitude of improvement over the free space decay lifetime of more than a week. We see that for sufficiently close distances $z_0$, the decay rate increases with $\omega$, spanning many orders of magnitude over a small frequency bandwidth. Furthermore, we see that with increasing distance $z_0$, the total decay rate is suppressed exponentially by the evanescent tail of the surface magnon. More specifically, we see in the exponential dependence $e^{-2q(\weg)z_0}$ that in order for rate enhancement to be effective, $z_0$ should be comparable to or ideally smaller than $1/q \sim d$. For a 200 nm film, enhancement begins to saturate for $z_0 < 20$ nm. In terms of a potential experiment, these are promising parameters which could result in a total transition rate of $10^4$ s$^{-1}$. Finally, we note that at distances $z_0$ extremely near to the surface, effects such as material losses or nonlocality may cause the behavior of the transition rate to deviate slightly from the predicted behavior.

Thinner films offer even more drastic capabilities for enhancement. The dipole transition rate and Purcell factor scale as $\eta^3$, which means that shrinking the film thickness $d$ even by conservative factors can result in a rapid increase in the maximum transition rate achievable. This $\eta^3$ scaling is exactly the same scaling found for Purcell factors of electric dipole transition enhancement in the vicinity of highly confined electrostatic modes such as SPPs \cite{rivera2016shrinking, caldwell2013low, rivera2017making}.

It is also worthwhile to consider not only the total transition rates, but also the Purcell factors. The right side axis of Figure~\ref{fig:dipole_rates}(b) shows the Purcell factor for spin relaxation into SMPs, computed as the ratio between the enhanced transition rate and the free space transition rate, and denoted as $F_p(\omega) = \Gamma_{\text{dipole}}/\Gamma_0$. We note that while the transition rate in the magnonic environment is technically the sum of the SMP emission rate and the radiative rate, in our systems the radiative rate is so small that it need not be considered.

Having established the duality between electric and magnetic surface polaritonics in the context of Purcell enhancement, other important conclusions about the scope and utility of SMPs follow. Most notably, Purcell factors for higher order magnetic processes should scale with mode confinement identically to those for the corresponding electric processes. Given an emitter-material system that can support such processes, it should be possible to compute transition rates of higher order processes such as magnetic quadrupole transitions and multi-magnon emission processes. Electromagnetic duality implies that a magnetic quadrupole transition Purcell factor, for instance, should scale as $\propto \eta^5$. For emission into modes confined to factors of 1000 or more, this enhancement factor could easily exceed $10^{15}$, eluding to the possibility of making highly forbidden magnetic quadrupole processes observable.

\subsection{Emission with in-plane anisotropy}

Thus far, we have considered geometries of MnF$_2$ in which the anisotropy axis of the crystal is out of the plane of a thin film (in the $z$ direction). Past work has brought both theoretical interest as well as experimental studies on antiferromagnetic surface interfaces in which the magnetic permeability anisotropy axis lies in-plane. In other words, the material has negative permeability in the out-of-plane direction as well as one in-plane direction, while having a permeability of 1 in the other in-plane direction. This geometry gives rise to an rich anisotropic dispersion relation of SMP modes, which in turn result in a nontrivial angular dependence for processes of spontaneous emission. We summarize those findings here.

 \begin{figure}[h!]
	\centering
	\includegraphics[width=0.4\textwidth]{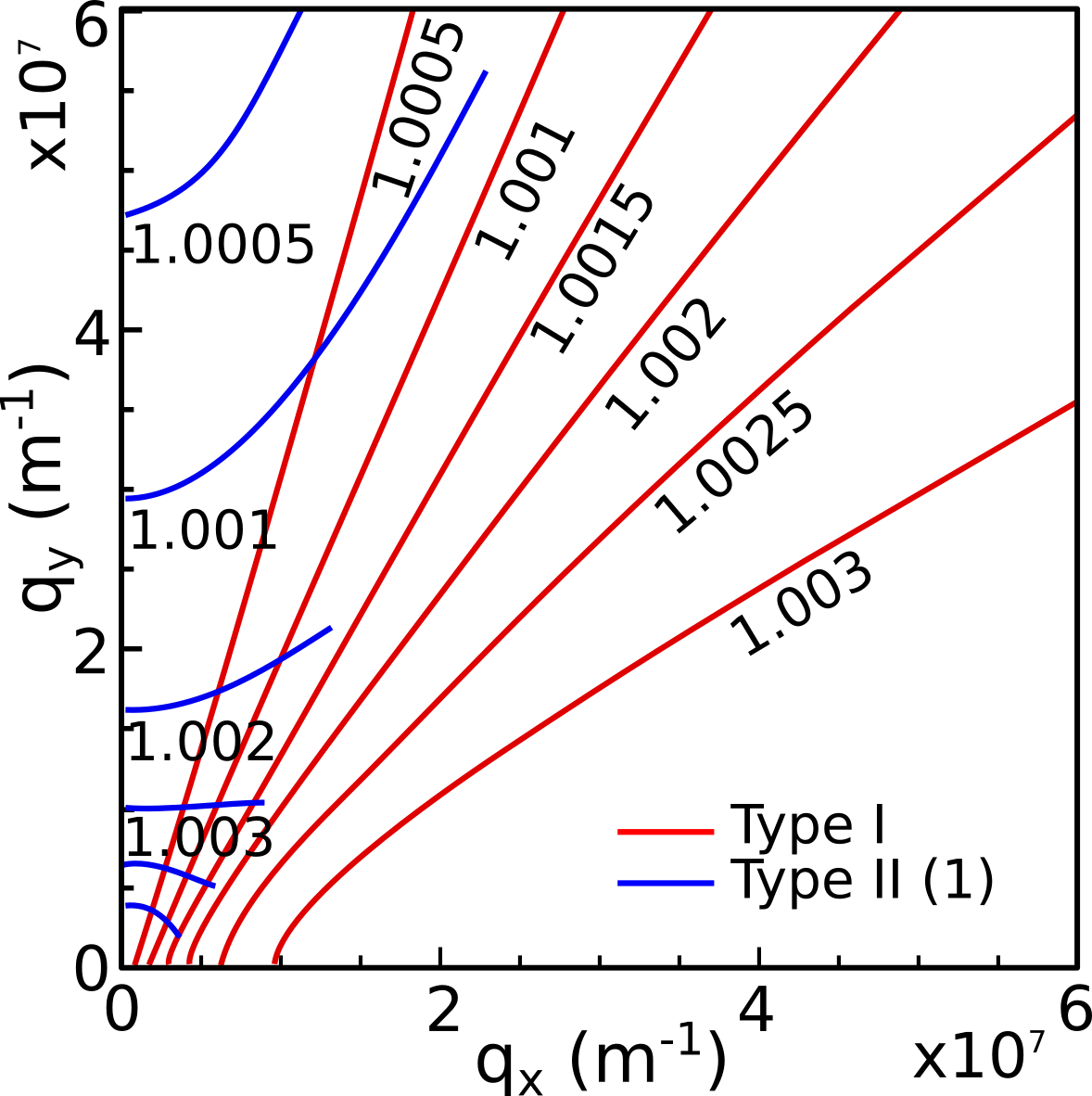}
	\caption{\textbf{Dispersion for anisotropic modes.} Isofrequency contours for MnF$_2$ of thickness $d=200$ nm. The frequency labels are given as $\omega/\omega_0$, where $\omega_0$ is the resonance frequency of the material. The first type I modes are shown in red, while the type II modes with $n=1$ are shown in blue.}
	\label{fig:anisotropic_isofreq}
\end{figure}

\begin{figure*}[!ht]
	\centering
	\includegraphics[width=\textwidth]{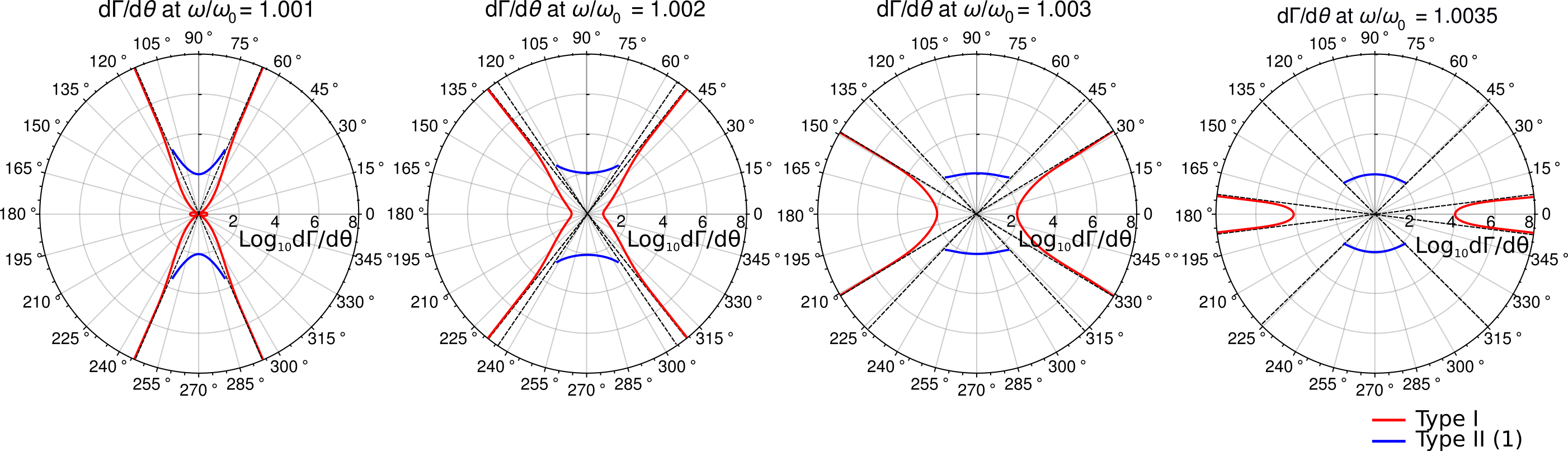}
	\caption{\textbf{Angular distribution of SMP emission.} Magnetic dipole transition rate per unit angle $d\Gamma^{(eg)}/d\theta$ for radiation into SMPs on a 200 nm thick slab of MnF$_2$. The radial axis shows $d\Gamma^{(eg)}/d\theta$ plotted on a log scale in units of s$^{-1}$. The first type I modes are shown in red and the first type II modes are shown in blue. Dashed lines indicate the angular cutoffs $\theta_x$ and $\theta_y$ for each type of mode. Note that at low frequencies $\theta_x$ and $\theta_y$ become very close. We additionally note that for $\omega/\omega_0 > 1.0035$, the type I mode branch shown in red vanishes entirely, leaving only the type II modes.}
	\label{fig:polar_dGammadtheta}
\end{figure*}

For the in-plane anisotropic geometry with $\mu = (\mu(\omega), 1, \mu(\omega))$, the dispersion (obtained again by solving Maxwell
's equations for a quasimagnetostatic scalar potential) is given by solutions to:
 \begin{equation}
	e^{qd\sqrt{\beta(\theta,\omega)}} = \frac{1 - \mu(\omega)\sqrt{\beta(\theta,\omega)}}{1 + \mu(\omega)\sqrt{\beta(\theta,\omega)}},
 \end{equation}
where $\beta(\theta,\omega) = \cos^2\theta+\sin^2\theta/\mu(\omega)$ and $\theta$ is the in-plane propagation angle measured with respect to the $x$-axis. When $\beta > 0$, the mode function has a $z$-dependence of $\cosh(qz)$ or $\sinh(qz)$, dependent on the parity of the solution. When $\beta < 0$, the modes have a $\cos(qz)$ or $\sin(qz)$ dependence. We note that the $\beta < 0$ solutions have a multiply branched structure which correspond to higher harmonic modes, just as with the in-plane isotropic case discussed throughout the text. Furthermore, recalling that $\mu < 0$ and examining $\beta(\theta,\omega)$, we see that for angles of propagation near 0, $\beta$ will be positive, while for angles of propagation near $\pi/2$, $\beta$ is negative. Based on the sign of $\beta$, we can classify the modes into two distinct types. We refer to $\beta > 0$ modes as type I modes, and $\beta < 0$ modes as type II modes. The fundamental type I modes propagate in the range $\theta \in (0,\theta_x)$, where $\theta_x = \tan^{-1}(\sqrt{-\mu(\omega)})$, while the type II modes with $n=1$ propagate in the range $\theta \in (\theta_y, \pi/2)$, with $\theta_y = \cos^{-1}(1/\sqrt{-\mu(\omega)})$. The angular propagation ranges for the type I modes and the lowest order type II mode are non-overlapping, and the gap between $\theta_x$ and $\theta_y$ increases with $\omega$.

The dispersion for even type I and type II modes are respectively given as:
 \begin{align}
 	q_{\text{I}} &= -\frac{1}{d\sqrt{\beta(\theta,\omega)}}\tanh^{-1}\left(\frac{1}{\mu(\omega)\sqrt{\beta(\theta,\omega)}}\right), \\
 	q_{\text{II}}^n &= \frac{1}{d\sqrt{-\beta(\theta,\omega)}}\tan^{-1}\left(\frac{1}{\mu(\omega)\sqrt{-\beta(\theta,\omega)}} + n\frac{\pi}{2}\right),
 \end{align}
where $n$ is an integer. We see that for even type I modes, only a single band of surface polariton modes exists, while for type II modes, a richer structure with harmonics exists due to the multivalued nature of the arctangent, just as in the in-plane isotropic case. In Figure~\ref{fig:anisotropic_isofreq}, we see the isofrequency contours for the dispersion in the case of in-plane anisotropy. We clearly observe that the mode structure is anisotropic, in that type I modes behave differently than type II modes. We comment briefly on the polarization of the modes. The in-slab $\mathbf{H}$-field polarization of the type I and II modes are respectively given as
 \begin{equation}
 	\hat{\varepsilon}_q = \begin{cases}
    \displaystyle\frac{\hat{q}\cosh(qz) + i\sinh(qz)\hat{z}}{\sqrt{2}}, & \text{type I} \\
 	\displaystyle\frac{\hat{q}\cos(qz) + i\sin(qz)\hat{z}}{\sqrt{2}}, & \text{type II}
    \end{cases}.
 \end{equation}

Applying the same formalism as before, the rate of emission into SMPs per unit angle by a $z$-oriented spin flip of strength $\mu_B$ is given by

\begin{equation}
 	\frac{d\Gamma^{(eg)}}{d\theta} = \frac{\mu_B^2 \mu_0 \weg}{2\pi\hbar}\frac{q^3(\theta,\weg)|\bm{\sigma}_{eg}\cdot\hat{\epsilon}_{\mathbf{q}}|^2}{C_q(\theta,\weg)|v_g(\theta,\weg)|}e^{-2q(\theta,\weg)z_0}.
\end{equation}

The total rate is as per usual obtained by integrating over all angles:
\begin{equation}
 	\Gamma^{(eg)} = \frac{\mu_B^2 \mu_0 \weg}{2\pi\hbar} \int_0^{2\pi} d\theta \frac{q^3(\theta,\weg)|\bm{\sigma}_{eg}\cdot\hat{\epsilon}_{\mathbf{q}}|^2}{C_q(\theta,\weg)|v_g(\theta,\weg)|}e^{-2q(\theta.\weg)z_0}.
\end{equation}

In Figure~\ref{fig:polar_dGammadtheta} we see the lossless differential decay rate $d\Gamma^{(eg)}/d\theta$ plotted as a function of polar angle $\theta$ for a $z$-oriented spin flip transition at different emitter frequencies $\omega$. We see that with increasing frequency, the angular spread of type I modes narrows, while the angular spread of type II modes increases. We can understand this behavior in terms of the availability and confinement of modes for different propagation angles $\theta$. The most highly confined modes are the type I modes near the angular cutoff. As $\omega$ increases the confinement of type I modes at low angles increases, while the confinement of type II modes decreases. This system exhibits the interesting property that tuning the frequency of the emitter over a narrow bandwidth dramatically shapes the angular spectrum of polariton emission. An interesting consequence is that inhomogeneous broadening of the emitter could play a strong role in determining the observed angular spectrum of magnons emitted.

\begin{figure}[h!]
	\centering
	\includegraphics[width=0.5\textwidth]{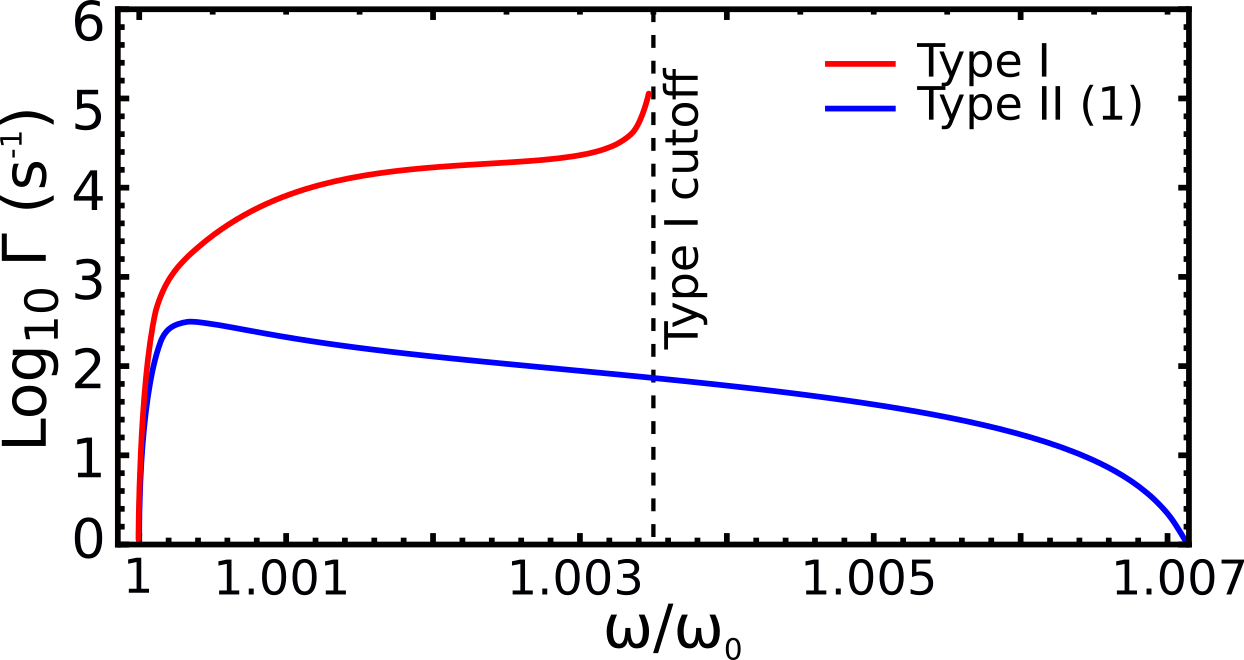}
	\caption{\textbf{Magnetic dipole transition rate for in-plane anisotropic MnF$_2$.} Magnetic dipole transition rate for a $z$-oriented dipole transition a distance $z_0=5$ nm from the surface into two different SMP modes in a $d=200$ nm thick anisotropic slab of MnF$_2$. The type I mode emits most strongly but over a narrower range of frequencies. The cutoff frequency is the frequency at which the first type I mode no longer satisfies the boundary conditions. The first order type II mode is emitted more weakly but is supported over the entire range of frequencies for which $\mu(\omega) < 0$.}
	\label{fig:anisotropic_rates}
\end{figure}

In Figure~\ref{fig:anisotropic_rates}, we see the total  transition rate $\Gamma^{(eg)}$ for a dipole emitter above MnF$_2$ oriented with the anisotropy axis in the $y$ direction. While the transition rates of both modes are greatly enhanced compared to the free space transition rate of order $10^{-6}$ $s^{-1}$, the type I mode benefits approximately two orders of magnitude more than the first type II mode. The Purcell factors for the type I mode in particular ranges from $10^{10}$ to $10^{12}$, and is thus quite comparable to Purcell factors obtained for the in-plane isotropic discussed previously. In this sense, we see that extreme enhancement of MD transition rates is achievable in both crystal orientations. The dispersion relation, however, is notably different in these cases. Further alterations to the dispersion in either geometry can be made using an external applied field, resulting in nonreciprocal propagation of modes. The net result is a highly flexible platform for ultrafast interaction between magnetic transitions and matter.


\section{Conclusions and Outlook}

We have shown that exceptionally confined surface magnon polaritons, such as those on antiferromagnetic materials, could speed up magnetic transitions by more than 10 orders of magnitude, bridging the inherent gap in decay rates which typically separates electric and magnetic processes. We predict that these confined magnetic surface modes in systems with realizable parameters may exhibit confinement factors in excess of $10^4$. We developed the theory of magnon polaritons and their interactions with emitters in a way that unifies this set of materials with other more well-known polaritonic materials, casting light on opportunities to use these materials to gain unprecedented control over spins in emitters. 

To push the field of magnon polaritonics at THz frequencies forward, it will be necessary to identify an ideal experimental platform for manipulating these modes and interfacing them with matter. As these modes exist at very low frequencies, experiments probing emitter interactions will need to take place under cryogenic conditions. Another question is what class of emitters may be well-suited to interact with these polaritonic modes. This is made challenging by the very narrow polaritonic bands of antiferromagnetic materials, as well as the few existing materials. This latter problem of course can be solved, as there are many more antiferromagnetic materials, which may support a negative permeability. Another interesting direction is the consideration of 2D antiferromagnetic materials. In terms of existing materials, a potential emitter system which can interact with these magnons is ErFeO$_3$, which has several electric and magnetic dipole transitions in the range between 0.25 and 1.5 THz \cite{mikhaylovskiy2017selective}. It could also prove interesting to consider GHz-THz orbital angular momentum transitions between high energy levels in Rydberg atoms. Processes involving the emission of multiple surface magnons, or mixed processes with the emission of a magnon polariton in addition to one or more excitations of another nearby material, could also be considered. In any case, surface magnon polaritons provide an interesting new degree of control over magnetic degrees of freedom in matter as well as a means to consider magnetic analogs at THz frequencies of many famous effects in plasmonics and polaritonics.


\newpage
\begin{acknowledgments}
The authors thank Charles Roques-Carmes and Nicolas Romeo for help reviewing the manuscript. Research supported as part of the Army Research Office through the Institute for Soldier Nanotechnologies under contract no. W911NF-18-2-0048 (photon management for developing nuclear-TPV and fuel-TPV mm-scale-systems). Also supported as part of the S3TEC, an Energy Frontier Research Center funded by the US Department of Energy under grant no. DE-SC0001299 (for fundamental photon transport related to solar TPVs and solar-TEs). I.K. is an Azrieli Fellow, supported by the Azrieli Foundation, and was partially supported by the Seventh Framework Programme of the European Research Council (FP7-Marie Curie IOF) under grant no. 328853-MC-BSiCS. N.R. recognizes the support of the DOE Computational Science Graduate Fellowship (CSGF) Number DE-FG02-97ER25308.
\end{acknowledgments}

\bibliography{magnonbib}

\end{document}